\documentclass[11pt,aps,showkeys,draft]{revtex4}
\usepackage{amssymb,amsfonts,amsmath}
\usepackage{bbm,bm}
\usepackage{times}
\newcommand{\p}{{\prime}}
\newcommand{\ra}{\rangle}
\newcommand{\la}{\langle}
\newcommand{\h}{\mathcal{H}}

\begin{document}

\title{Generalized Schmidt decomposition based on injective tensor norm}

\author{Levon Tamaryan}
\affiliation{Physics Department, Yerevan State University,
Yerevan, 375025, Armenia}
\author{DaeKil Park}
\affiliation{Department of Physics, Kyungnam University, Masan,
631-701, Korea}
\author{Sayatnova Tamaryan}
\affiliation{Theory Department, Yerevan Physics Institute,
Yerevan, 375036, Armenia}

\begin{abstract}
We present a generalized Schmidt decomposition for a pure system
with any number of two-level subsystems. The basis is symmetric
under the permutation of the parties and is derived from the
product state defining the injective tensor norm of the state. The
largest coefficient quantifies the quantum correlation of the
state. Other coefficients have a lot of information such as the
unentangled particles as well as
the particles whose reduced states are completely mixed. The
decomposition clearly distinguishes the states entangled in
inequivalent ways and have an information on the applicability to
the teleportation and superdense coding when the given quantum
state is used as a quantum channel.
\end{abstract}


\maketitle

\section{Introduction}

The Schmidt decomposition for bipartite systems\cite{smid} is a
very important tool in quantum information and quantum computing
theories. It shows whether two given states are related by a local
unitary transformation or not\cite{ekert}, which states are
applicable for perfect teleportation\cite{bentel} and superdense
coding\cite{dence}, and  whether it is possible to transform a
given bipartite pure state to another pure state by local
operations and classical communications\cite{niels}. Many
substantial results have been obtained with the help of the
Schmidt normal form and its generalization to the multipartite
states is a task of prime importance\cite{acin,ben-schm}.

Carteret {\it et al} developed a method for such a generalization
for pure states of an n-party system, where the dimensions of the
individual state spaces are finite but otherwise
arbitrary\cite{hig}. The idea of this method is the following.
First, one finds the product vector that gives maximal overlap with
a given state function. Next, one considers product vectors whose
constituents are orthogonal to the constituents of the first
product vector and finds among them the product vector that gives
maximal overlap with the state function. Continuing in this way,
one finds a set of orthonormal product states and presents the
state function as a linear combination of these product vectors.
Since the first product vector is a stationarity point, the
resulting canonical form contains a minimal set of state
parameters.

Surprisingly, an arbitrary multipartite quantum state has many
canonical forms. The reason is that the stationarity
equations(SEQ) defining stationarity points are nonlinear
equations and in general have many solutions of different
types\cite{wei}. For example, in the case of generic three qubit
states there are six product vectors that are solutions of
SEQ\cite{shared}. Moreover, some highly entangled states are
surrounded by the one-parametric set of equally distant product
states that have the same maximal overlap with the state
function\cite{tri}. All of these product vectors are equally good
in a sense that all resulting canonical forms use minimal number
of product vectors from a factorizable orthonormal basis.
Therefore, it is necessary to make a physically motivated choice of
the product vector that yields the best factorizable basis. More
precisely, the problem is the following. Each of bases suggests
its own set of invariants which are coefficients of the expansion
of the state function. And the task is to specify the set of
invariants that gives an effective description of the quantum
state.

Our approach is based on the injective tensor norm of quantum
states\cite{wern}. The overlap of a given state with any complete
product state attains maximums at several points. These maximums
are eigenvalues and corresponding product states are eigenvectors
of SEQ. The injective tensor norm takes the value of one of these
maximums and thus makes a unique choice of the eigenvector. This
choice classifies pure quantum states by types as follows.
Consider three qubit pure states. There are six eigenvectors and
eigenvalues of SEQ and therefore the injective tensor norm can
have six different expressions. Since it is a single valued
function, each of expressions should has its own range of
definition in which they are deemed applicable. Consequently, each
of solutions of SEQ has its own applicable domain depending on
state parameters in which they give the relevant eigenvalue. As a
result, the space of the state parameters is split into six
regions and each region has its own specific eigenvector defining
the injective tensor norm. Each of three qubit pure states should
belong to the one of six regions of the parameter space except
states that lie on joint surfaces separating different regions.
This classification provides a geometric picture of entangled
regions.


The main objective of this work is to construct the generalized
Schmidt decomposition(GSD) for generic three qubit states. We
analyze the complete set of SEQ solutions, specify the relevant
solution for a given state and calculate GSD coefficients
explicitly in terms of the state parameters. In this way we obtain
a new set of invariants of three qubit states. Since there are
different sets of local invariants\cite{acin,sud}, one may wonder
what is the advantage of the obtained coefficients. For this
purpose we study the properties of the constructed GSD.

The expansion coefficients exhibit the physically significant
properties of pure states. The largest coefficient $g$ is the
injective tensor norm of the state. It is a very useful quantity
and defines some entanglement measures \cite{vedr, Shim, barn,
wei, bno}. Apart from that, we will present a rigorous proof that
if a three qubit pure state has a completely mixed one qubit
reduced state, then $g^2=1/2$. This sheds some new light on the
applicability of the quantum state to the teleportation and
superdense coding. There is a conjecture that the criterion for
perfect quantum teleportation is $g^2=1/2$\cite{tele}. Now this
conjecture can be reformulated as follows: a pure state can be
used for perfect teleportation and superdense coding if it has a
completely mixed state as a reduced state. The other expansion
coefficient, {\it say} $h$, has an information on the presence or
absence of an unentangled particle in a given quantum state. We
will show in the following that $h=0$ is  a separability criterion
for pure states of a general multi-qubit system \cite{per}.

This paper is organized as follows. In Sec. II we specify GSD for
multiqubit systems. In Sec. III we analyze general properties of
GSD in the case of three qubit systems. In Sec. IV we construct
GSD for three qubit W-type states. In Sec. V we construct GSD for
a class of multiqubit W-type states. In Sec. VI we construct GSD
for Greenberger-Horne-Zeilinger(GHZ)-type states. In Sec. VII we
make concluding remarks.

\section{Generalized Schmidt decomposition}

Consider $n$-partite pure systems with the Hilbert space
$\h=\h_1\otimes\h_2\otimes\cdots\otimes\h_n$. The injective tensor
norm $g(\psi)$ of a given $n$-partite pure state $|\psi\ra$ is
defined as
\begin{equation}
g(\psi)=\sup|\la\chi_1\chi_2\cdots\chi_n|\psi\ra|,
\end{equation}
\noindent where the supremum is over all tuples of vectors
$|\chi_k\ra\in \h_k$ with $\|\chi_k\|=1$ \cite{wern}. The nearest
product state $|q\ra=|q_1\ q_2\cdots q_n\ra$ must satisfy
stationarity equations \cite{wei,kobes}
\begin{equation}\label{gen.near}
\la q_1q_2\cdots \widehat{q_k}\cdots q_n|\psi\ra=g|q_k\ra, \quad
k=1,2,\cdots n
\end{equation}
\noindent where the caret means exclusion. These equations can be regarded as
nonlinear eigenvalue problem: all eigenvalues $g$ and eigenvectors
$|q\ra$ satisfy it for a given state function $|\psi\ra$. In
general, Eq.(\ref{gen.near}) has several different solutions.
Hereafter we consider only the solution corresponding to the
eigenvalue of largest value. This solution, known as dominant
eigenvector, defines the injective tensor norm of a given quantum
state.


Consider now $n$-qubit system. For each single-qubit state
$|q_k\ra$ there is, up to arbitrary phase, a unique single-qubit
state $|p_k\ra$ orthogonal to it. From these single-qubit states
$|q_k\ra$ and $|p_k\ra$ one can form a set of $2^n$ $n$-qubit
product states which form a basis in the full Hilbert space $\h$.
Any vector $|\psi\ra\in \h$ can be written as a linear combination
of vectors in the set. Then from SEQ (\ref{gen.near}) it follows
that all the coefficients of the product states $|q_1 \cdots
q_{k-1}p_k q_{k+1}\cdots q_n\ra(k=1,2,\cdots n)$ are zero. Thus
any pure state can be written in terms of $2^n-n$ product states.
Furthermore, the phases of vectors $|p_k\ra$ are free and we can
choose them so that all the coefficients $t_k$ of vectors
$|p_1\cdots p_{k-1}q_kp_{k+1}\cdots p_n\ra(k=1,2,\cdots n)$ be
positive. Still we have a freedom to make a phase shift
$|p_k\ra\rightarrow e^{2i\pi/(n-1)}|p_k\ra$ which remains
unchanged $t_k$ and $g$. We use this freedom to vary the phase
$\varphi$ of the component $e^{i\varphi}h|p_1p_2\cdots p_n\ra$
($h\geq0$ is understood) within the interval
$-\pi/(n-1)\leq\varphi\leq\pi/(n-1)$.

Thus the decomposition has $n+1$ real and $2^n-2n-1$ complex
parameters. After taking into account the normalization condition,
one can show that $2^{n+1}-3n-2$ real numbers parameterize the
sets of inequivalent pure states \cite{number}.

\smallskip

{\bf Theorem\;1.} The $k$th qubit is completely unentangled if and
only if $h(\psi)=0$ and $t_i(\psi)=0$ for $i\neq k$.

{\bf Proof.} Suppose first qubit is completely unentangled and its
state vector is $|q_1\ra$. We have
$|\psi\ra=|q_1\ra\otimes|\psi^\p\ra$. Let the product state
$|q_2q_3\cdots q_n\ra$ be the nearest state of $|\psi^\p\ra$. Then
GSD of $|\psi^\p\ra$ takes the form
\begin{equation}\label{theor}
|\psi\p\ra=g^\p|q_2q_3\cdots q_n\ra+\sum_{i=2}^n t_i^\p|p_2\cdots
p_{i-1}q_ip_{i+1}\cdots
p_n\ra+\cdots+e^{i\varphi^\p}h^\p|p_2p_3\cdots p_n\ra.
\end{equation}
\noindent Since the nearest state of the state $|\psi\ra$ is, up
to a phase, the product state $|q_1q_2\cdots q_n\ra$, then
$g(\psi)=g^\p,\;h(\psi)=0,\;t_1(\psi)=h^\p$ and $t_i=0,
i=2,3...n$. The inverse is also true. From $h(\psi)=0$ and
$t_i(\psi)=0$ for $i\neq1$ it follows that all the terms in GSD
which do not contain $|q_1\ra$ vanish and
$|\psi\ra=|q_1\ra\otimes|\psi^\p\ra$. Similarly, theorem is true
if any other qubit is unentangled.

\smallskip

 Consider now $n=2$ and $n=3$ cases. For simplicity we will use
notations $|0_i\ra$ and $|1_i\ra$ for vectors $|q_i\ra$ and
$|p_i\ra$ respectively. Also we will omit sub-indices $i$ whenever
it does not create misunderstanding. In the case of two qubit
states the expansion reduces to the Schmidt decomposition
$|\psi\ra=g|00\ra+h|11\ra$ with $g\geq h\geq0$. Consider
three-qubit case.

\section{Three qubit states.}

Decomposition takes the form
\begin{equation}\label{3.gsd}
|\psi\ra=g|000\ra+t_1|011\ra+t_2|101\ra + t_3|110\ra+
e^{i\varphi}h|111\ra.
\end{equation}
The coefficients should satisfy conditions
\begin{equation}\label{3.cond}
g\geq\max(t_1,t_2,t_3,h),\;t_1\geq0,\;t_2\geq0,\;t_3\geq0,\;h\geq0,\;
-\frac{\pi}{2}\leq\varphi\leq\frac{\pi}{2}.
\end{equation}
These conditions do not specify GSD uniquely. Eq.(\ref{3.gsd}) is
the GSD normal form of the state $|\psi\ra$ if and only if $g$ is
the injective norm of the state $|\psi\ra$. There are highly
entangled states which can be written in a form of
Eq.(\ref{3.gsd}) in six different bases. One basis, where the
largest coefficient is injective tensor norm of the state and
corresponding product state is the dominant eigenvector of SEQ,
gives true GSD.  The others, where the largest coefficient is not
injective tensor norm and corresponding product states are
eigenvectors but are not dominant eigenvectors, do not. The
example with W-type states, which is given in the next section,
illustrates this more clearly.

Now we formulate a theorem which shows whether a given three qubit
pure state has a completely mixed one qubit reduced state or not.
The proof of Theorem\;2 uses a particular lower bound on $g$ given
in the next section. In this reason we present the proof in
Appendix.

\smallskip

{\bf Theorem\;2.} One-particle reduced state of $k$th qubit is
completely mixed if and only if $$t_k=0\quad {\rm and}\quad
g^2=1/2.$$

Consider now several interesting examples.

\section{W-type states}

Our first example that we shall discuss in detail is a family of
four-parametric W-type states \cite{w}
\begin{equation}\label{t.psi}
|\psi\ra=a|100\ra+b|010\ra+c|001\ra+d|111\ra.
\end{equation}
Without loss of generality we consider only the case of positive
parameters $a,b,c,d$.

\subsection{Solutions of stationarity equations}

Stationarity equations (\ref{gen.near}) have six different
solutions for each of these states\cite{shared}. Four solutions
are simple and represent four terms on the right-hand side of
Eq.(\ref{t.psi})
\begin{subequations}\label{t.lar}
\begin{equation}\label{t.lara}
|q_1q_2q_3\ra=|100\ra,\quad g=a;
\end{equation}
\begin{equation}\label{t.larb}
|q_1q_2q_3\ra=|010\ra,\quad g=b;
\end{equation}
\begin{equation}\label{t.larc}
|q_1q_2q_3\ra=|001\ra,\quad g=c;
\end{equation}
\begin{equation}\label{t.lard}
|q_1q_2q_3\ra=|111\ra,\quad g=d.
\end{equation}
\end{subequations}
These four solutions owe their existence to the fact that the
states Eq.(\ref{t.psi}) have a hidden symmetry. Namely, there
exist local unitary transformations that interchange the positions
of any of two coefficients in Eq.(\ref{t.psi}). Due to this
symmetry any entanglement measure is invariant under the
permutations of all of the state parameters $a,b,c,d$.

The fifth nontrivial solution is
\begin{eqnarray}\label{t.near}
 & &|q_1\ra=\frac{\sqrt{r_ar_d}|0_1\ra +
\sqrt{r_br_c}|1_1\ra}{4S\sqrt{ad+bc}},\quad
|q_2\ra=\frac{\sqrt{r_br_d}|0_2\ra +
\sqrt{r_ar_c}|1_2\ra}{4S\sqrt{ac+bd}}, \\\nonumber
 & &|q_3\ra=\frac{\sqrt{r_cr_d}|0_3\ra +
\sqrt{r_ar_b}|1_3\ra}{4S\sqrt{ab+cd}},\quad g=\frac{L}{S},
\end{eqnarray}
where
$$r_a=a(b^2+c^2+d^2-a^2)+2bcd,\quad r_b=b(a^2+c^2+d^2-b^2)+2acd,
\quad r_c=c(a^2+b^2+d^2-c^2)+2abd,$$
$$r_d=d(a^2+b^2+c^2-d^2)+2abc,\quad L=\sqrt{(ab+cd)(ac+bd)(ad+bc)}$$
and $S$ is the area of the cyclic quadrangle with sides
$(a,b,c,d)$. Note that the quadrangle is not unique as the sides
can be arranged in different orders. But all these quadrangles
have the same area given by Heron's formula
$S=\sqrt{(s-a)(s-b)(s-c)(s-d)}$, where $s=(a+b+c+d)/2$ is
semiperimeter.

The sixth solution is obtained from the solution (\ref{t.near}) by
the interchange
\begin{equation}\label{t.int}
d\rightarrow-d,\quad r_d\rightarrow-r_d.
\end{equation}
Due to symmetry one can make one of the following interchanges
instead
\begin{eqnarray*}
 & &a\rightarrow-a,\quad r_a\rightarrow-r_a;\\
 & &b\rightarrow-b,\quad r_b\rightarrow-r_b;\\
 & &c\rightarrow-c,\quad r_c\rightarrow-r_c
\end{eqnarray*}
and obtain the same solution. However, it does not mean that any
of parameters becomes negative. All the parameters are positive
and the interchange (\ref{t.int}) shows the existence of a pair of
dual solutions.

\subsection{Dominant eigenvector}

We have six different product states $|q_1q_2q_3\ra$ and therefore
we can form six sets of factorizable bases from vectors
$|q_1\ra,|q_2\ra,|q_3\ra$ and $|p_1\ra,|p_2\ra,|p_3\ra$. Since all
these product states satisfy SEQ, all the expansions of the state
function $|\psi\ra$ in these bases give the canonical form
Eq.(\ref{3.gsd}). Therefore there are six factorizable bases where
the state function Eq.(\ref{t.psi}) has a canonical form
Eq.(\ref{3.gsd}). The question at issue is whether there are
effective expansions or not. If it is indeed the case, then what
are the effective expansions which have physically meaningful
coefficients?

The injective tensor norm $g$ has six different expressions. Since
it is a single valued function, each expression has its own
applicable domain depending on state parameters and these
applicable domains are split up by separating surfaces.
Consequently each of solutions (\ref{t.lar}) and (\ref{t.near})
has its own applicable domain depending on state parameters in
which they are dominant eigenvectors. Thus four-dimensional sphere
given by normalization condition $a^2+b^2+c^2+d^2=1$ is split into
six parts and each part has its own dominant eigenvector.

The above classification suggests a natural choice of a
factorizable basis. We choose the product state $|q_1q_2q_3\ra$
that is dominant eigenvector in a given part of sphere. Next we
form product states from constituents $|q_i\ra$ and orthogonal to
them vectors $|p_i\ra$. Thus it is necessary to separate the
validity domains and to make clear which of the solutions should
be applied for a given state. It is a nontrivial task and for a
thorough analysis we refer to Ref.\cite{shared}. Here we present
some necessary results.

In highly entangled region parameters $(a,b,c,d)$ form a cyclic
quadrilateral, injective tensor norm is expressed in terms of the
the circumradius of the quadrangle and corresponding solution is
Eq.(\ref{t.near}). In slightly entangled region injective tensor
norm is the largest coefficient and corresponding solutions are
Eq.(\ref{t.lar}). Also there are states in between for which both
formulae are valid. These states, called second type shared
quantum states, separate slightly and highly entangled states and
can be ascribed to both types. Another specific states, called
first type shared quantum states, are those for which injective
tensor norm is a constant and is defined by $g^2=1/2$. These
states allow perfect quantum teleportation and superdense coding
scenario.

\subsection{Highly entangled W-type states}

Highly entangled region is defined by inequalities
\begin{equation}\label{t.rarb}
r_a\geq0,\quad r_b\geq0,\quad r_c\geq0,\quad r_d\geq0.
\end{equation}
The constituents of the dominant eigenvector are given by
Eq.(\ref{t.near}) and the task is to present the state function as
a linear combination of product vectors
\begin{equation}\label{t.gsd}
|\psi\ra=g|q_1q_2q_3\ra+t_1|q_1p_2p_3\ra+t_2|p_1q_2p_3\ra+
t_3|p_1p_2q_3\ra+e^{i\varphi}h|p_1p_2p_3\ra,
\end{equation}
where $\la p_i|q_i\ra=0,\;i=1,2,3$.

The calculation of the coefficients requires advanced mathematical
technique. One has to factorize polynomials of degree ten. We
would like to suggest a simple way. First one convinces oneself
that each factor below is a root for the corresponding polynomial
and next finds the proportionality coefficient in some particular
case. The derivation of $h$ is the most complicated out of all
coefficients and one can use the hint: if $a=b+c+d$, then
$r_b=r_c=r_d=-r_a$. The resulting answer is
\begin{equation}\label{t.fin}
g=\frac{L}{2S},\; t_1=\frac{Lr_1}{4S(ad+bc)},\;
t_2=\frac{Lr_2}{4S(bd+ac)},\; t_3=\frac{Lr_3}{4S(cd+ab)},\;
\varphi=\frac{\pi}{2},\; h=\frac{\sqrt{r_ar_br_cr_d}}{4LS},
\end{equation}
\noindent where $r_1,\;r_2,\;r_3$ are the lengths of the Bloch
vectors of first, second and third qubits
\begin{equation}\label{t.ri}
r_1=|b^2+c^2-a^2-d^2|,\; r_2=|a^2+c^2-b^2-d^2|,\;
r_3=|a^2+b^2-c^2-d^2|.
\end{equation}
In fact, this set gives a fruitful description of the state. The
invariant $g$ is expressed in terms of the circumradius of the
cyclic quadrangle $a,b,c,d$ and gives geometric and Groverian
entanglement measures of the state. First type shared states are
defined by $r_1r_2r_3=0$ which is to say that the reduced state of
one of particles is completely mixed. It is quite obvious that one
of coefficients $t_i$ must vanish for these states. On the other
hand if $r_k=0$, then $g^2=1/2$ and the corresponding state allows
teleportation(and dense coding) scenario. For perfect
teleportation the receiver should choose $k$th particle at initial
stage in order to perform the task. Thus the coefficients $t_i$
contain an information on the applicability to the teleportation
and precisely indicate which particle the receiver should choose.
Second type shared states lie on the separating surface and this
surface is defined by $r_ar_br_cr_d=0$, i.e $h=0$. We conclude
that $h>0$ for highly entangled states and $h=0$ for second type
shared states.

To complete the analysis let us consider the remaining slightly
entangled cases.

\subsection{Slightly entangled W-type states}

At least one of quantities $r_a,r_b,r_c$ and $r_d$ should be
negative in this case. Note that two of them can not be negative
simultaneously. Indeed,
\begin{equation}\label{s.non}
br_a+ar_b=2(ac+bd)(bc+ad)\geq0
\end{equation}
and therefore $r_a$ and $r_b$ can not be negative together.
Similarly, any pair of quantities $r_a,r_b,r_c,r_d$ can not be
negative simultaneously. Thus one of them should be negative and
others should be positive and there are four possibilities.
Consider these cases in turn.

\paragraph{Solution (\ref{t.lara}).} The first case is
\begin{equation}\label{s.1}
r_a\leq0.
\end{equation}
The dominant eigenvector is given by (\ref{t.lara}). In order to
obtain GSD one has to simply relabel the basis of the first qubit
$|0_1\ra\leftrightarrow|1_1\ra$. Then the final GSD coefficients
are
\begin{equation}\label{s.coef1}
g=a,\quad h=0,\quad t_1=d,\quad t_2=c,\quad t_3=b.
\end{equation}

\paragraph{Solution (\ref{t.larb}).} The second case is
\begin{equation}\label{s.2}
r_b\leq0.
\end{equation}
The dominant eigenvector is given by (\ref{t.larb}). Now one has
to relabel the basis of the second qubit
$|0_2\ra\leftrightarrow|1_2\ra$. GSD coefficients are
\begin{equation}\label{s.coef2}
g=b,\quad h=0,\quad t_1=c,\quad t_2=d,\quad t_3=a.
\end{equation}

\paragraph{Solution (\ref{t.larc}).} The third case is
\begin{equation}\label{s.3}
r_c\leq0.
\end{equation}
The dominant eigenvector is given by (\ref{t.larc}) and one has to
relabel the basis of the third qubit
$|0_3\ra\leftrightarrow|1_3\ra$. GSD coefficients are
\begin{equation}\label{s.coef3}
g=c,\quad h=0,\quad t_1=b,\quad t_2=a,\quad t_3=d.
\end{equation}

\paragraph{Solution (\ref{t.lard}).} The fourth case is
\begin{equation}\label{s.4}
r_d\leq0.
\end{equation}
The dominant eigenvector is given by (\ref{t.lard}) and one has to
relabel the bases of all qubits $|0_1\ra\leftrightarrow|1_1\ra,\;
|0_2\ra\leftrightarrow|1_2\ra,\;|0_3\ra\leftrightarrow|1_3\ra$.
Resulting GSD coefficients are
\begin{equation}\label{s.coef4}
g=d,\quad h=0,\quad t_1=a,\quad t_2=b,\quad t_3=c.
\end{equation}

All of four cases can be summarized as follows. If the state
Eq.(\ref{t.psi}) is slightly entangled, then

i) its last coefficient $h=0$ vanishes and $g$ takes the value of
the largest coefficient
\begin{equation}\label{s.larg}
h\equiv 0,\quad g=\max(a,b,c,d)
\end{equation}

ii) GSD takes the form
\begin{equation}\label{s.gsd}
|\psi\ra=g|000\ra+t_1|011\ra+t_2|101\ra + t_3|110\ra,
\end{equation}

iii) All the inequalities (\ref{s.1}), (\ref{s.2}), (\ref{s.3})
and (\ref{s.4}) become the same lower bound on $g$
\begin{equation}\label{s.low}
g^2\geq t_1^2+t_2^2+t_3^2+2\frac{t_1t_2t_3}{g}.
\end{equation}

The obvious conclusion is that $h\neq0$ only for the highly
entangled states and identically vanishes for the slightly
entangled states.

Note that four slightly entangled cases (\ref{s.1}), (\ref{s.2}),
(\ref{s.3}) and (\ref{s.4}) together with the highly entangled
case (\ref{t.rarb}) cover the fourdimensional sphere given by the
normalization condition. Thus the sixth solution does not create
its own factorizable basis. However, it does not mean at all that
the solution is unphysical. The state Eq.(\ref{t.psi}) is a four
parametric state whiles generic three qubit states should have six
parameters. Thus the state Eq.(\ref{t.psi}) is not generic and the
applicable domain of the sixth's solution has been shrunk to the
separating surface $r_1r_2r_3=0$ in this case. It is easy to check
that the sixth solution results the same GSD coefficients
Eq.(\ref{t.fin}) for first type shared states. But it will create
its own GSD for generic states.

Let's now examine multiqubit W-states.

\section{Multiqubit W-type states}
In this section we consider one-parametric n-qubit W-states
\begin{equation}\label{n.psi}
|\psi\ra=a\left(|100\cdots0\ra+|0100\cdots0\ra+\cdots+
|00\cdots010\ra\right)+b|00\cdots01\ra.
\end{equation}
Slightly entangled region is given by $r_n=(n-1)a^2-b^2<0$
\cite{toward}. In this region the last product state $|0 \cdots
01\rangle$ is the nearest separable state and
\begin{equation}\label{n.sl}
g=b,\quad h=0.
\end{equation}
In highly entangled region $r_n>0$ and, consequently,
$S_n=(n-1)^2a^2-b^2>0$. The constituent states for the closest
separable states are respectively
\begin{equation}\label{n.near}
|q_1\ra=\cdots=|q_{n-1}\ra=\frac{a\sqrt{(n-1)(n-2)}|0\ra+
\sqrt{r_n}|1\ra} {\sqrt{S_n}},\quad
|q_n\ra=\frac{\sqrt{(n-1)r_n}|0\ra+b\sqrt{n-2}|1\ra}{\sqrt{S_n}}.
\end{equation}
Straightforward calculation gives
\begin{equation}\label{n.fin}
g=(1-b^2)^{\gamma+1/2}\left[\frac{n-2}{S_n}
\right]^\gamma,\;t_n=\sqrt{(n-2)r_n} \left[ \frac{r_n}{S_n}
\right]^\gamma,\;h=b\sqrt{n-1} \left[ \frac{r_n}{S_n}
\right]^\gamma,\;\varphi=\frac{\pi}{n-1},
\end{equation}
where
$$\gamma=\frac{n-2}{2}.$$

Expressions (\ref{n.fin}) have the same meanings as in  the
three-qubit case. First, $r_n=0$ forces $g^2=1/2$. Second,
$g^2>1/2$ and $h=0$ means the state is slightly entangled. Third,
$g^2<1/2$ and $h=0$ means $b=0$ and, therefore, the last qubit is
unentangled. Fourth, we conjecture that: all the states with
$r_n=0$ allow the teleportation scenario and the receiver should
choose $n$th qubit. In summary, suggested GSD indicates the
applicability to the teleportation and distinguishes the
unentangled particles as well as completely entangled particles.

\section{GHZ-type states.}

Consider now the extended GHZ state \cite{ghz}
\begin{equation}\label{ghz.psi}
|\psi\ra=a|000\ra+b|001\ra+c|110\ra+d|111\ra.
\end{equation}
Stationarity equations have two solutions\cite{tri}
\begin{subequations}\label{ghz.sol}
\begin{equation}\label{ghz.sol1}
|q_1\ra=|0\ra,\quad|q_2\ra=|0\ra,\quad
|q_3\ra=\frac{a|0\ra+b|1\ra}{\sqrt{a^2+b^2}},\quad
g=\sqrt{a^2+b^2};
\end{equation}
\begin{equation}\label{ghz.sol2}
|q_1\ra=|1\ra,\quad|q_2\ra=|1\ra,\quad
|q_3\ra=\frac{c|0\ra+d|1\ra}{\sqrt{c^2+d^2}},\quad
g=\sqrt{c^2+d^2}.
\end{equation}
\end{subequations}
The state function takes the canonical form Eq.(\ref{3.gsd}) in
two bases. In both cases nonzero coefficients of the decomposition
can be presented as follows
\begin{equation}\label{ghz.c}
g=\max\left(\sqrt{a^2+b^2},\sqrt{c^2+d^2}\right),\quad
t_3=\frac{ac+bd}{g}, \quad h=\frac{|ad-bc|}{g}.
\end{equation}
This set of GSD coefficients describes the extended GHZ-type
states almost in the same way as bipartite systems. Since
$g^2\geq1/2$, there is no highly entangled region for GHZ-type
states. In this sense W-state is more entangled than GHZ-state.
When the extended GHZ-state is most entangled, i.e. $g^2=1/2$, it
is applicable for both teleportation and dense coding and the
situation is same in the case of bipartite systems. In contrast to
W-type case, there is no region where $h$ is identically zero.
Only on condition $ad=bc$ the canonical coordinate $h$ vanishes.
Thus if $h$ vanishes, then the state is biseparable and again the
same is true for two-qubit systems. The only difference from
two-qubit case is that there is an extra term with the coefficient
$t_3$. It shows that the third particle is unentangled when $h=0$.

\section{Summary}

We have used a general way of constructing generalized Schmidt
decomposition proposed in Ref.\cite{hig} and specified a unique
decomposition for arbitrary composite systems consisting of
two-level subsystems. This decomposition suggests a new set of
local invariants. Three qubit pure states have following six
invariants
$$g,\;t_1,\;t_2,\;t_3,\;h,\;\varphi.$$
We have considered general three-qubit and W-type $n$-qubit
systems whose injective tensor norms were already derived
analytically. All the invariants have been calculated explicitly
and expressed in terms of state parameters $a,b,c,d$. It is shown
that they provide a profound information on the quantum states.
The largest coefficient $g$ gives two entanglement measures and
together with the last coefficient $h$ clearly distinguishes the
states entangled in inequivalent ways. Namely, for W-type states
there is a region where the function $h(a,b,c,d)$ is identically
zero and there is no such region for GHZ-type states. Furthermore,
isolated zeros of the function $h(a,b,c,d)$ indicate the
appearance of the unentangled particles. The coefficients $t_i$
are related to Bloch vectors and reveal the existence of
completely mixed reduced states. Thereby they show whether or not
a given state is applicable for perfect teleportation(and dense
coding) and precisely indicate which particle the receiver should
choose at initial stage in order to perform the task.

We have derived a lower bound on injective tensor norm
Eq.(\ref{s.low}) for slightly entangled states(h=0). A natural
question now arises: is it possible to derive such a lower bound
for arbitrary three qubit states? This question is closely related
to the following problem. For every measure, including geometric,
there must be a set of states which are maximally entangled, or at
least approaching some lower bound. And the problem is to find
these states explicitly. If one derives a strong lower bound on
$g$, then one can see when the lower bound is saturated and obtain
maximally entangled states.

A lower bound on $g$ for generic three qubit states can be derived
by making use of GSD form (\ref{3.gsd}). Indeed, consider a state
$|\psi_{\rm rest}\ra=
t_1|011\ra+t_2|101\ra+t_3|110\ra+e^{i\varphi}h|111\ra$. Suppose
$|pr\ra$ is its nearest product state and $\mu(t_1,t_2,t_3,h)$ is
its injective tensor norm. Obviously
$$g\geq|\la pr|\psi\ra|=|g\la pr|000\ra+\mu|.$$
This is a generic lower bound and its derivation reduces to the
derivation of $|pr\ra$. Unfortunately, known methods do not allow
us to solve SEQ for $|\psi_{\rm rest}\ra$ and find $|pr\ra$. Since
the phases $\varphi$ can be absorbed in the definitions of basis
vectors, $|\psi_{\rm rest}\ra$ is a four parametric state. It is
interesting to note that the states of type $|\psi_{\rm rest}\ra$
are the only remaining four parametric states whose injective norm
has not been obtained analytically so far. There is a good reason
to calculate this injective tensor norm and obtain a lower bound
on it for three qubit states.

\smallskip

\begin{acknowledgments}
ST thanks Tzu-Chieh Wei for useful discussions.

This work was supported by the Kyungnam University Foundation,
2008.
\end{acknowledgments}

\appendix
\section{Proof of Theorem\;2}

{\bf Proof.} Suppose $t_1=0$ and  $g^2=1/2$. Equation
(\ref{3.gsd}) allows to express the Bloch vectors of qubits in
terms of GSD coefficients. Denote by $r_1$ the norm of the Bloch
vector of the first particle. Straightforward calculation yields
\begin{equation}\label{3.mod}
r_1^2=4h^2t_1^2+\left(g^2+t_1^2-t_2^2-t_3^2-h^2\right)^2.
\end{equation}
Then $t_1=0$ and  $g^2=1/2$ together with the normalization
condition $g^2+t_1^2+t_2^2+t_3^2+h^2=1$ force $r_1=0$. Therefore
the reduced state of the first particle is completely mixed.\\
Suppose now that the reduced state of the first particle is a
completely mixed state. Then $r_1$ should vanish which is to say
that
\begin{equation}\label{3.mix}
ht_1=0,\quad g^2+t_1^2-t_2^2-t_3^2-h^2=0.
\end{equation}
Either $t_1=0$ or $h=0$. If $h=0$, then the state Eq.(\ref{3.gsd})
can be transformed to the state Eq.(\ref{t.psi}) by a local
unitary transformation $|0_i\ra\leftrightarrow|1_i\ra$. In this
reason both states have the same GSD and Eq.(\ref{3.gsd})
coincides with Eq.(\ref{s.gsd}). Thus the inequality
Eq.(\ref{s.low}) holds for both states. On the other hand $h=0$
forces $g^2=t_2^2+t_3^2-t_1^2$ and this is contradicted by
inequality Eq.(\ref{s.low}). Consequently $t_1=0$ and the the
normalization condition forces $g^2=1/2$. This ends the proof of
the theorem.

\begin{thebibliography}{99}
\bibitem{smid}E. Schmidt, {\it Zur theorie der linearen und
nichtlinearen integralgleighungen}, Math. Ann. {\bf 63},
433(1907).
\bibitem{ekert}A. Ekert and P. L. Knight,
{\it Entangled quantum systems and the Schmidt decomposition}, Am.
J. Phys. {\bf 63}, 415 (1995).
\bibitem{bentel}C. H. Bennett, G. Brassard, C. Cr\'epeau, R. Jozsa,
A. Peres, and W. K. Wootters, {\it Teleporting an unknown quantum
state via dual classical and Einstein-Podolsky-Rosen channels},
Phys. Rev. Lett. {\bf 70}, 1895(1993).
\bibitem{dence}C. H. Bennett and S. J. Wiesner,
{\it Communication via one- and two-particle operators on
Einstein-Podolsky-Rosen states}, Phys. Rev. Lett. {\bf 69},
2881(1992).
\bibitem{niels}M. A. Nielsen, {\it Conditions for a Class
of Entanglement Transformations}, Phys. Rev. Lett. {\bf 83},
436(1999).
\bibitem{acin}A. Ac\'in, A. Andrianov, L. Costa, E. Jan\'e,
J. I. Latorre, and R. Tarrach, {\it Generalized Schmidt
decomposition and classification of three-quantum-bit states},
Phys. Rev. Lett. {\bf 85}, 1560(2000).
\bibitem{ben-schm}C. H. Bennett, S. Popescu, D. Rohrlich, J. A.
Smolin, and A. V. Thapliyal, {\it Exact and asymptotic measures of
multipartite pure-state entanglement}, Phys. Rev. A {\bf63},
012307(2000).
\bibitem{hig}H. A. Carteret, A. Higuchi, and A. Sudbery,
{\it Multipartite generalisation of the Schmidt decomposition}, J.
Math. Phys. {\bf 41}, 7932(2000).
\bibitem{wei}T.-C. Wei and P. M. Goldbart,
{\it Geometric measure of entanglement and applications to
bipartite and multipartite quantum states}, Phys. Rev. A {\bf 68},
042307(2003).
\bibitem{shared}L. Tamaryan, D. K. Park, J.-W Son, and S. Tamaryan,
{\it Geometric measure of entanglement and shared quantum states},
Phys. Rev. A {\bf 78}, 032304(2008).
\bibitem{tri}L. Tamaryan, D. K. Park and S. Tamaryan,
{\it Analytic expressions for geometric measure of three qubit
states}, Phys. Rev. A {\bf77}, 022325(2008).
\bibitem{wern}R. Werner and A. Holevo,
{\it Counterexample to an additivity conjecture for output purity
of quantum channels}, J. Math. Phys. {\bf 43}, 4353(2002).
\bibitem{sud}A. Sudbery, {\it On local invariants of pure three-qubit
states}, J. Phys. A {\bf34}, 643(2001).
\bibitem{vedr}V. Vedral, M. B. Plenio, M. A. Rippin, and P. L. Knight,
{\it Quantifying Entanglement}, Phys. Rev. Lett. {\bf 78},
2275(1997).
\bibitem{Shim}A. Shimony,
{\it Degree of entanglement}, Ann. NY. Acad. Sci {\bf 755},
675(1995).
\bibitem{barn}H. Barnum and N. Linden,
{\it Monotones and invariants for multi-particle quantum states},
J. Phys. A: Math. Gen. {\bf 34}, 6787(2001).
\bibitem{bno}O. Biham, M. A. Nielsen and T. J. Osborne,
{\it An entanglement monotone derived from Grover's algorithm},
Phys. Rev. A {\bf65}, 062312(2002).
\bibitem{tele}E. Jung, M.-R. Hwang, D.K. Park, J.-W. Son,
and S. Tamaryan, {\it Perfect quantum teleportation and suerdense
coding with $P_{max}=1/2$ states}, arXiv:0711.3520v1[quant-ph].
\bibitem{per}A. Peres,
{\it Separability criterion for density matrices}, Phys. Rev.
Lett. {\bf 77}, 1413(1996).
\bibitem{kobes}D. Ostapchuk, G. Passante, R. Kobes, and G.
Kunstatter, {\it Geometric measures of entanglement and the
Schmidt decomposition}, arXiv:0707.4020v2[quant-ph].
\bibitem{number}N. Linden and S. Popescu,
{\it On multi-particle entanglement}, Fortschr. Phys. {\bf 46},
567(1998).
\bibitem{w}W. D\"{ur}, G. Vidal and J. I. Cirac,
{\it Three qubits can be entangled in two inequivalent ways},
Phys.Rev. A {\bf 62}, 062314(2000).
\bibitem{toward}L. Tamaryan, H. Kim, E. Jung,
M.-R. Hwang, D.K. Park, and S. Tamaryan, {\it Toward an
understanding of entanglement for generalized n-qubit W-states},
arXiv:0806.1314v1[quant-ph].
\bibitem{ghz}D. Greenberger, M. Horne, and A. Zeilinger,
{\it Going beyond Bell's theorem}, arXiv:0712.0921[quant-ph].
062317(2008).
\end{thebibliography}
\end{document}